\documentclass{mn2e}

\usepackage{psfig,subfigure,mncite}

\newcommand{\bvec}[1]{\mbox{\boldmath ${#1}$}}
\newcommand{\deriv}[2]{\mbox{${{\displaystyle d#1}\over {\displaystyle d#2}}$}}

\newcommand{\etal}{\mbox{\em et\ al.\ }}

\begin{document}
 

\title[Magnetic topology of AB Doradus]
{The global magnetic topology of AB Doradus}

\author
[M. Jardine, A. Collier Cameron \& J.-F. Donati]
{M. Jardine$^1$, 
\thanks{E-mail: moira.jardine@st-and.ac.uk}
A. Collier Cameron$^1$ \& J.-F. Donati$^2$ 
\\
$^1$School of Physics and Astronomy, Univ.\ of St~Andrews, 
St~Andrews, 
Scotland KY16 9SS \\
$^2$Laboratoire d¹Astrophysique, Observatoire Midi-Pyr\'en\'ees, 14 
Av. E. Belin, F-31400 Toulouse, France \\
\\
} 

\date{Received; accepted }

\maketitle

\begin{abstract}
 We have used Zeeman-Doppler maps of the surface field of the young,
 rapid rotator AB Dor (P$_{\rm rot}$ = 0.514 days) to extrapolate the
 coronal field, assuming it to be potential.  We find that the
 topology of the large-scale field is very similar in all three years
 for which we have images.  The corona divides cleanly into regions of
 open and closed field.  The open field originates in two mid-latitude
 regions of opposite polarity separated by about 180$^{\circ}$ of
 longitude.  The closed field region forms a torus extending almost
 over each pole, with an axis that runs through these two longitudes. 
 We have investigated the effect on the global topology of different
 forms of flux in the unobservable hemisphere and in the dark polar
 spot where the Zeeman signal is suppressed.  The flux distribution in
 the unobservable hemisphere affects only the low latitude topology,
 whereas the imposition of a unidirectional polar field forces the
 polar cap to be open.  This contradicts observations that suggest
 that the closed field corona extends to high latitudes and leads us
 to propose that the polar cap may be composed of multipolar regions.
 
\end{abstract}


\section{Introduction}
\begin{figure*}

  \begin{tabular}{cc}

   \psfig{figure=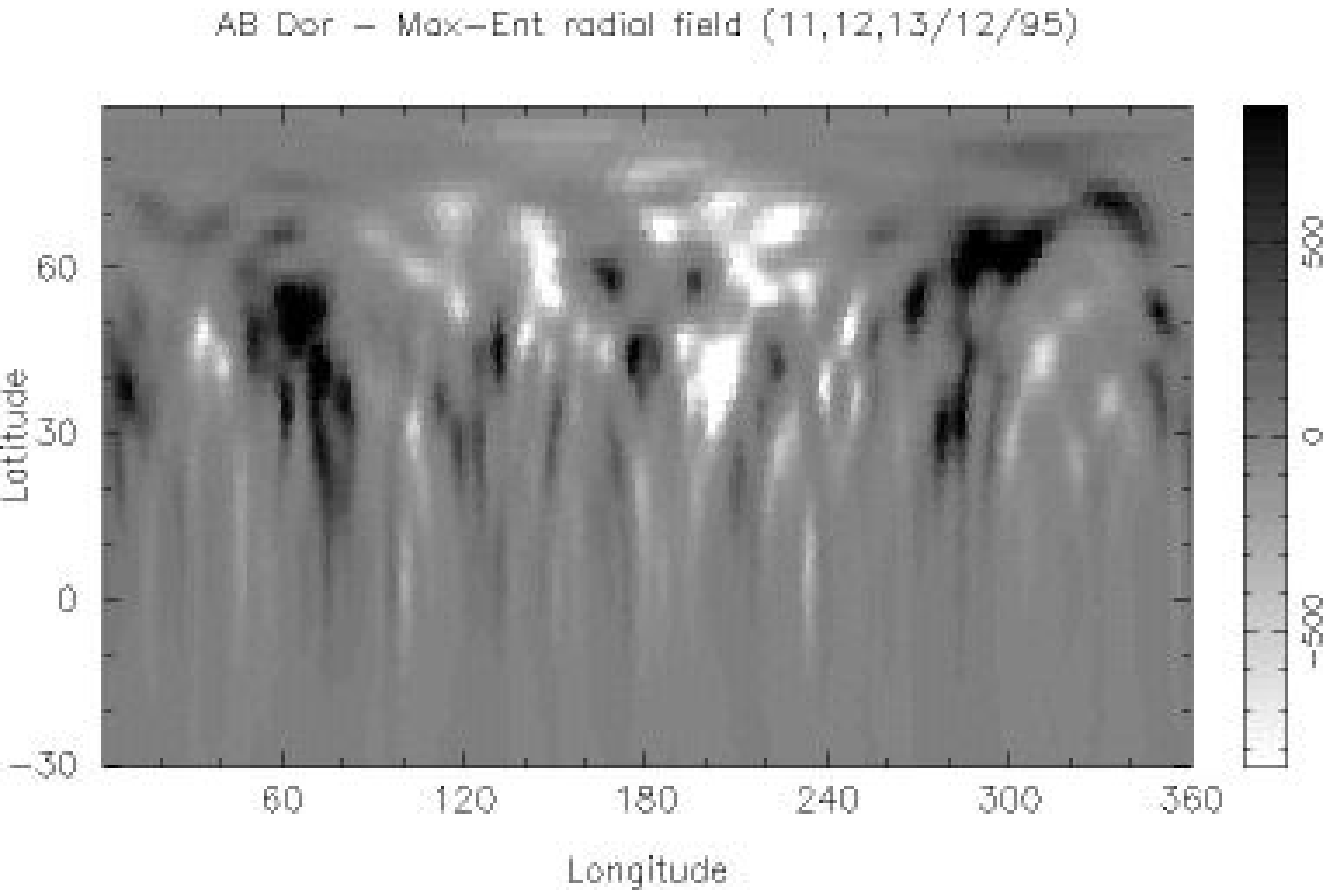,width=8cm} &
   \psfig{figure=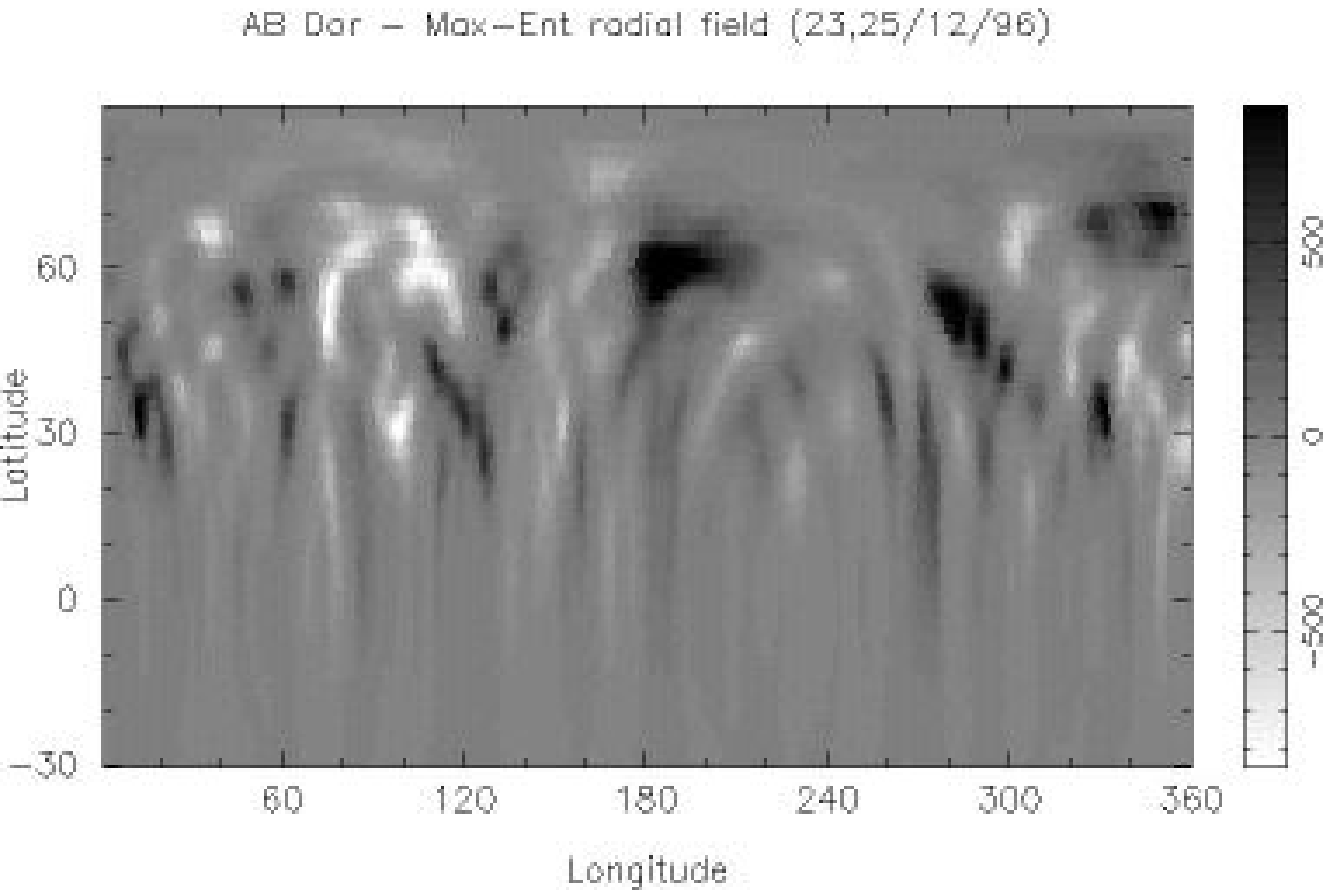,width=8cm}

  \end{tabular}

 \caption{Zeeman-Doppler images of the surface radial field of AB Dor
 on 1995 Dec 11-13 (top) and 1996 Dec 23-25 (bottom).  The scale bar 
 on the right is in Gauss.  Note
 that because the star is inclined at 60$^{\circ}$ to the observer,
 there is limited information in the lower hemisphere.}

  \label{zdi}
 
\end{figure*}
\begin{figure*}
	\def\subfigtopskip{4pt}
	\def\subfigbottomskip{4pt}
	\def\subfigcapskip{2pt}
	\centering
	\begin{tabular}{ccc}
    	\subfigure[]{
			\label{onehemis95_18_surf} 			
			\psfig{figure=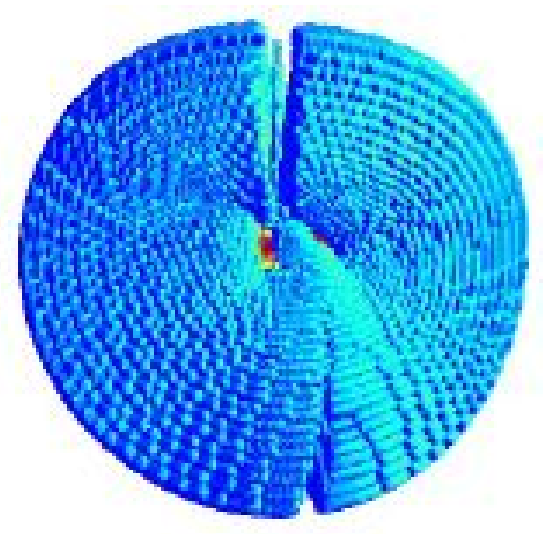,width=5cm}
			} &
		\subfigure[]{
			\label{onehemis95_18_closed} 						
			\psfig{figure=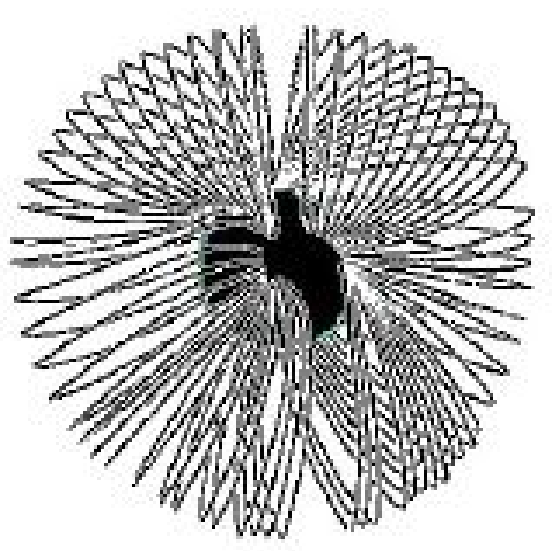,width=5cm}
			} &
		\subfigure[]{
			\label{onehemis95_18_open} 
			\psfig{figure=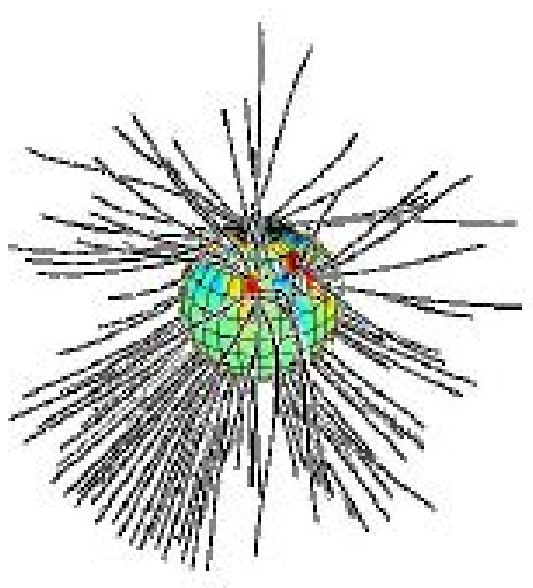,width=5cm}
			} \\
		\subfigure[]{
			\label{onehemis95_288_surf} 					
			\psfig{figure=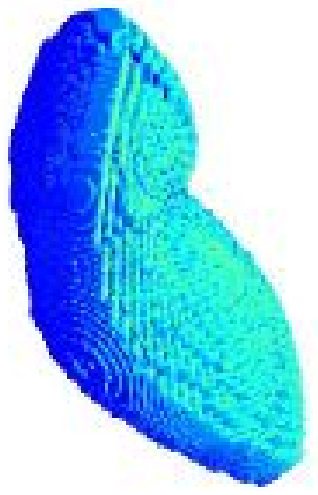,width=5.5cm}
			} &
		\subfigure[]{
			\label{onehemis95_288_closed} 		
			\psfig{figure=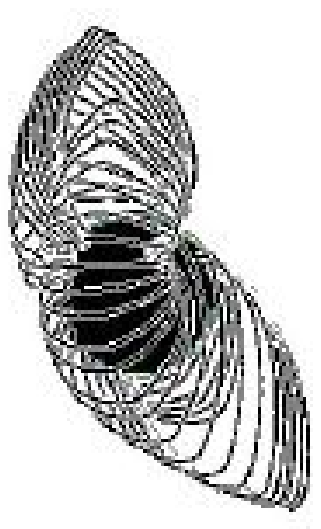,width=5cm}
			} &
		\subfigure[]{
			\label{onehemis95_288_open} 
			\psfig{figure=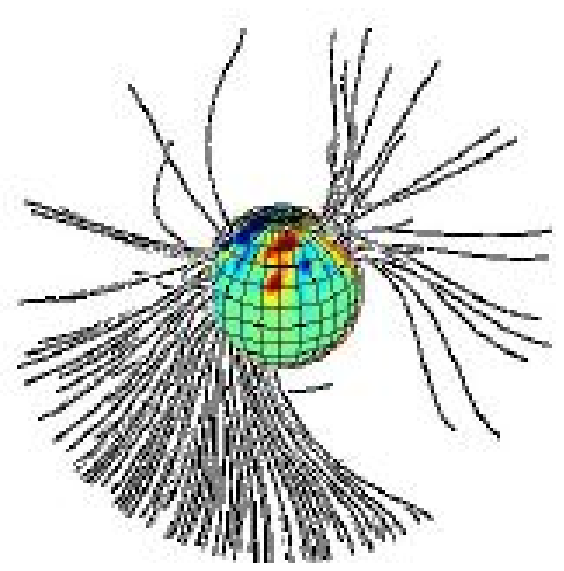,width=5cm}
			}  \\
	\end{tabular} 
	\caption[]{The global topology of the magnetic field of AB Dor for 
	the 1995 data set. The top three panels show a view from longitude 
	18$^{\circ}$ while the bottom panels show a view from longitude 
	288$^{\circ}$. Panels \ref{onehemis95_18_surf} and 
	\ref{onehemis95_288_surf} show the separator surfaces that separate the 
	closed and open field regions. Note that the gap in the surface at 
	the pole is due simply to problems in tracing field lines that pass 
	directly over the pole. Panels \ref{onehemis95_18_closed} and 
	\ref{onehemis95_288_closed} show the closed field lines just inside 
	the separator surface and panels \ref{onehemis95_18_open} and 
	\ref{onehemis95_288_open} show the open field lines just outside it. 
	In panels \ref{onehemis95_18_open} and 
	\ref{onehemis95_288_open} the surface radial map has been painted 
	onto the stellar surface. }
    \label{onehemis95}
\end{figure*}	
\begin{figure*}

  \begin{tabular}{cc}

   \psfig{figure=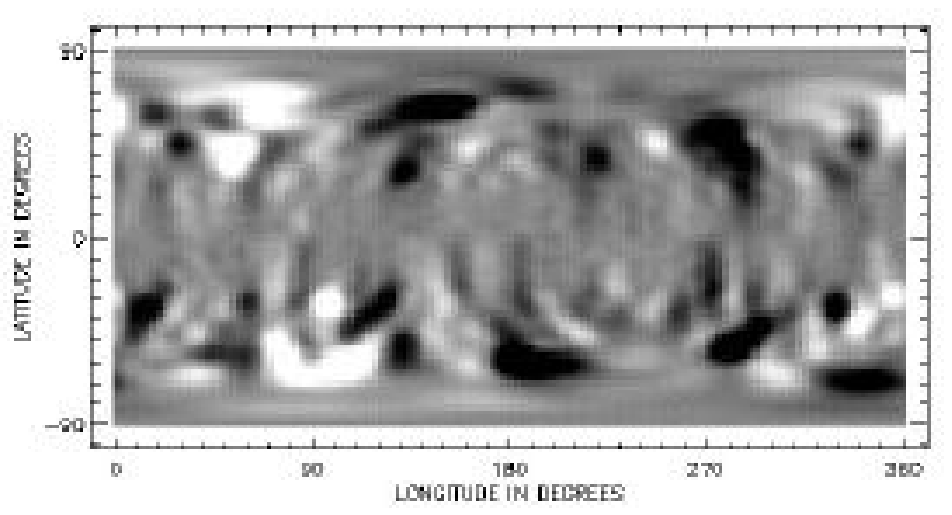,width=8cm}  &
   \psfig{figure=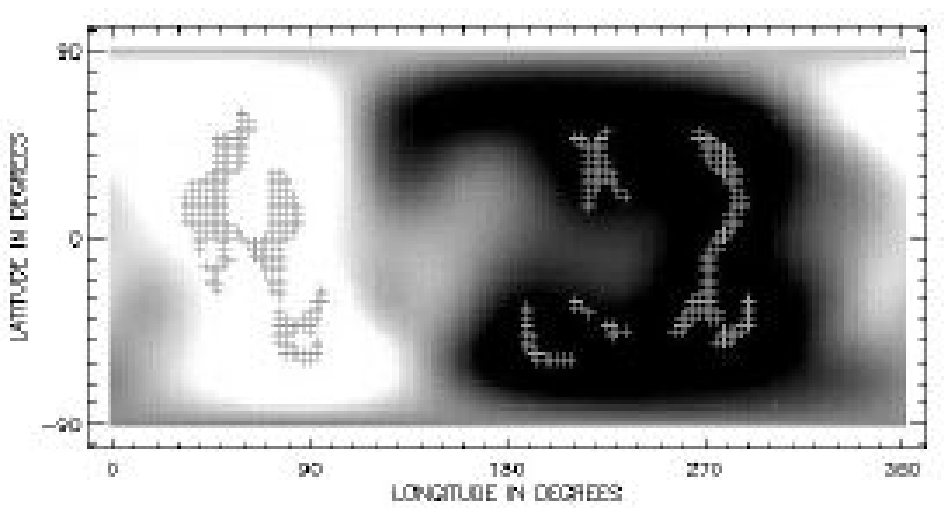,width=8cm}  \\
 
   \psfig{figure=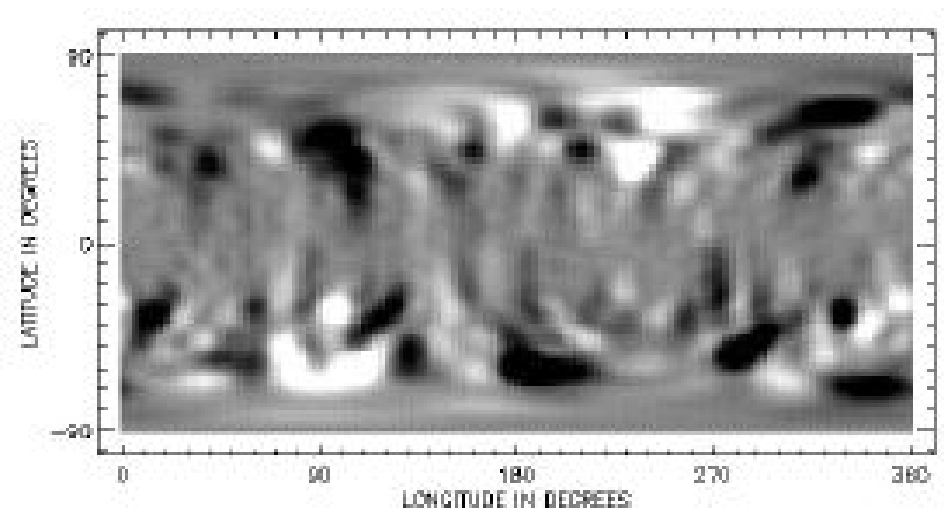,width=8cm}  &
   \psfig{figure=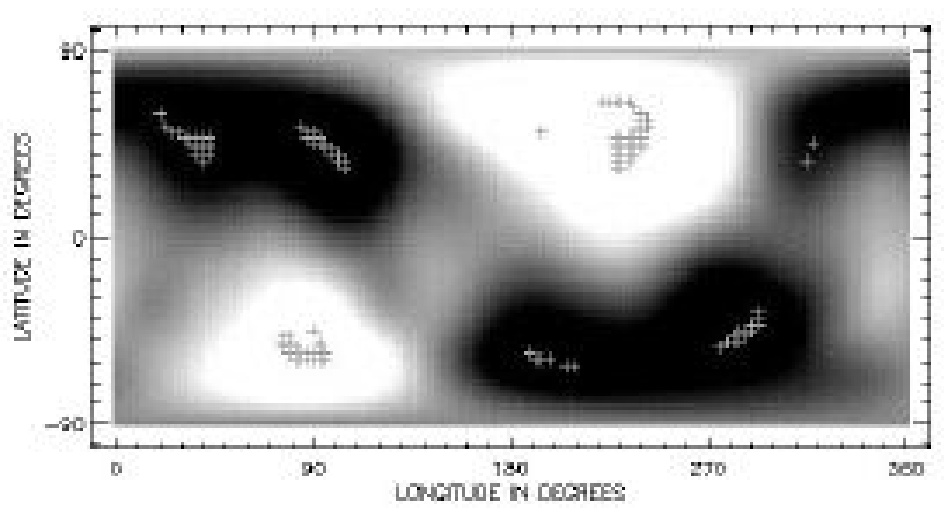,width=8cm}  \\
  
  \end{tabular}

 \caption{Images of the combined radial field with the 1995 map in the upper 
 hemisphere and the 1996 map in the lower hemisphere. The top two 
 panels show the case where the two maps are aligned such that both
hemispheres are symmetric about the equator (positive polarity open
region at the same longitude in each hemisphere). The lower two 
panels show an antisymmetric alignment (positive
polarity open regions 180 degrees apart in longitude in the two
hemispheres). Left-hand side panels show the surface radial field 
(greyscale extends from black at +500G to white at -500G), 
while panels on the right show the radial field at a height of 
0.6R$_{\star}$ above the surface (greyscale extends from black at +30G 
to white at  -30G). 
The crosses mark the  
positions of the footpoints of the open field lines.}

  \label{icont}
 
\end{figure*}

AB Dor is one of the most comprehensively observed of the known young
rapid rotators.  Long term studies of its photometric as well as its
X-ray variability are available \cite{amado01}.  It shows a small
(5-13 \%) rotational modulation in its X-ray emission \cite{kurster97}
and a large emission measure of $10^{52-53}$cm$^{-3}$ \cite{vilhu2001}
consistent with a very extended or a very dense corona.  The
observation of large coronal prominences trapped in co-rotation with
the star between 3 and 5 stellar radii from the rotation axis
\cite{cameron89cloud,cameron89eject,donati97abdor95,donati99abdor96}
suggest that the corona maintains a complex structure out to large
distances.

Several observations also suggest the presence of closed loops at high
latitudes.  Radio observations taken over a 6 month period
\cite{lim94} have shown two prominent peaks in the emission, separated
by 180$^{\circ}$ of longitude.  Lim {\em et al} suggest a model for
directed radio emission orginating from latitudes around 60$^{\circ}$. 
X-ray emission from such high latitudes is also believed to be
responsible for the flare observed with BeppoSAX \cite{maggio2000}
which showed no rotational modulation although the observations
spanned more than a whole rotation period.  Since the loop height
derived from modelling the flare decay phase was only 0.3R$_{\star}$,
Maggio {\em et al} claim that this flaring loop structure must have
been situated above 60$^{\circ}$ latitude where it would remain in
view throughout a rotation cycle.

These indicators of magnetic loops at high latitude are consistent
with Doppler images of AB Dor (and many other stars - see Strassmeier
1996) which show dark spots at or near the pole in addition to spots
at low latitudes.  Indeed, Zeeman-Doppler images (see Fig. 
\ref{zdi}) show flux at all latitudes on AB Dor (Donati \&
Collier Cameron 1997, Donati et al.  1999) {\em except} close to the
pole where the Zeeman signal is suppressed due to the low surface
brightness there.  What the Zeeman-Doppler images do show at the 
boundary of the dark polar cap is a ring of unidirectional azimuthal 
field. \scite{pointer01evol} suggest that this may be the result of the 
star's differential rotation dragging out meridional field lines at 
the edge of the polar cap to form an azimuthal ring.  In this case, 
the polarity of the field in the azimuthal ring would depend on the 
polarity of the field in the polar cap.  In particular, the fact that 
this ring is of uniform polarity, would suggest that the field in the polar 
cap is also of one polarity. There is one significant problem with 
this scenario, however. If we place a  unipolar field region at the 
pole, then some fraction of those field lines will be forced open by 
the pressure of the plasma they contain. If this ``polar hole'' 
extends down too far in latitude, then it will not be possible to 
explain the BeppoSAX observations that suggest that the closed corona 
extends to latitudes above 60$^{\circ}$.

In this paper we use the radial magnetic field maps shown in Fig. 
\ref{zdi} to extrapolate the coronal field.  Our aim is to
study the topology of the large scale field and to examine whether it
is possible to reconcile the observations indicating a closed corona
at high latitudes with the presence of a large polar spot.

\section{Extrapolating the coronal field}

We write the magnetic field $\bvec{B}$ in term of a flux function $\Psi$
such that $\bvec{B} = -\bvec{\nabla} \Psi$ and the condition that the
field is potential ($\bvec{\nabla}\times\bvec{B} =0$) is satisfied
automatically. The condition that the field is divergence-free then
reduces to Laplace's equation $\bvec{\nabla}^2 \Psi=0$. A solution in terms
of spherical harmonics can then be found:
\begin{equation}
 \Psi = \sum_{l=1}^{N}\sum_{m=-l}^{l} [a_{lm}r^l + b_{lm}r^{-(l+1)}]
         P_{lm}(\theta) e^{i m \phi},
\label{psifull}
\end{equation}
where the associated Legendre functions are denoted by $P_{lm}$. This then gives
\begin{equation}
B_r  =  -\sum_{l=1}^{N}\sum_{m=-l}^{l}
               [la_{lm}r^{l-1} - (l+1)b_{lm}r^{-(l+2)}]
               P_{lm}(\theta) e^{i m \phi}
\end{equation}
\begin{equation}
B_\theta  =  -\sum_{l=1}^{N}\sum_{m=-l}^{l} 
               [a_{lm}r^{l-1} + b_{lm}r^{-(l+2)}]
               \deriv{}{\theta}P_{lm}(\theta) e^{i m \phi}
\end{equation} 
\begin{equation} 
B_\phi  =  -\sum_{l=1}^{N}\sum_{m=-l}^{l} 
               [a_{lm}r^{l-1} + b_{lm}r^{-(l+2)}]
               \frac{P_{lm}(\theta)}{\sin \theta} ime^{i m \phi}.
\label{psi}
\end{equation}
The coefficients $a_{lm}$ and $b_{lm}$ are determined by imposing the
radial field at the surface and by assuming that at some height $R_s$
above the surface the field becomes radial and hence $B_\theta (R_s) =
0$ \cite{altschuler69}.  Since large slingshot prominences are
observed on AB Dor mainly around the co-rotation radius which lies at
$2.7R_\star$ from the rotation axis, we know that much of the corona
is closed out to those heights and so we set the value of $R_s$ to
$3.4R_\star$.  In order to calculate the field we used a code
originally developed by van Ballegooijen \etal (1998). 
\nocite{vanballegooijen98}

We use three data sets, obtained on 1995 Dec 11-13, 1996 Dec 23-25 and
1998 January 10-15 (see Fig.  \ref{zdi}).  In each case we extrapolate
the coronal field and distinguish between those field lines that are
closed and those that are open.  Fig. \ref{onehemis95} shows some
sample open field lines and also the {\em separator} surfaces that
separate the regions of open and closed field.  In all three cases the
global topology is similar and so we show only the results for the 
1995 data set. There are two dominant regions of open 
field lines formed at mid-latitudes about 180$^{\circ}$ of longitude 
apart.  They are of opposite polarity and hence form large helmet 
streamers, below which the field is predominantly closed.

\section{The effect of the unobservable hemisphere}
\begin{figure*}
	\def\subfigtopskip{4pt}
	\def\subfigbottomskip{4pt}
	\def\subfigcapskip{2pt}
	\centering
	\begin{tabular}{ccc}
    	\subfigure[]{
			\label{mixed0_220_surf} 			
			\psfig{figure=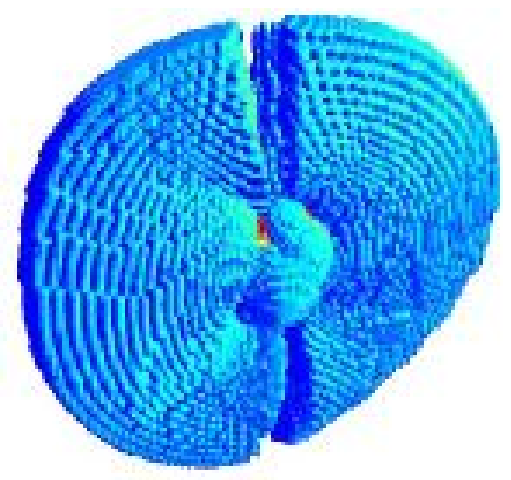,width=5cm}
			} &
		\subfigure[]{
			\label{mixed0_220_closed} 						
			\psfig{figure=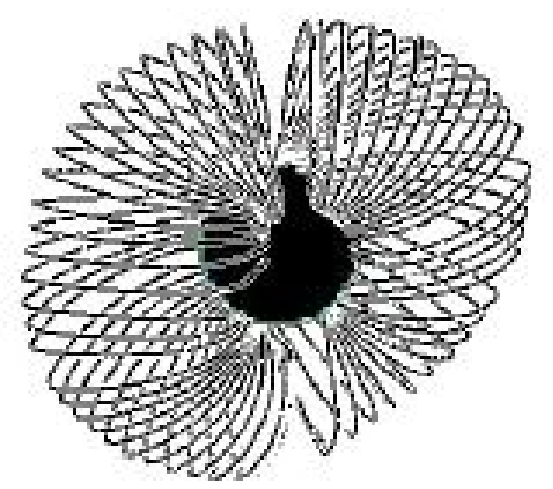,width=5cm}
			} &
		\subfigure[]{
			\label{mixed0_220_open} 
			\psfig{figure=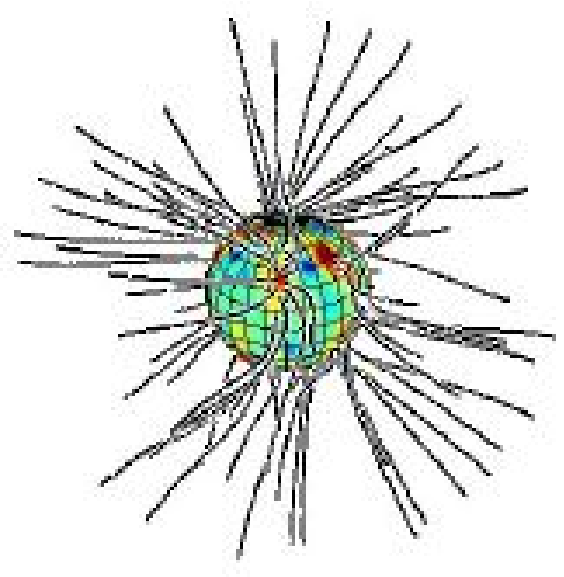,width=5cm}
			} \\
		\subfigure[]{
			\label{mixed0_130_surf} 					
			\psfig{figure=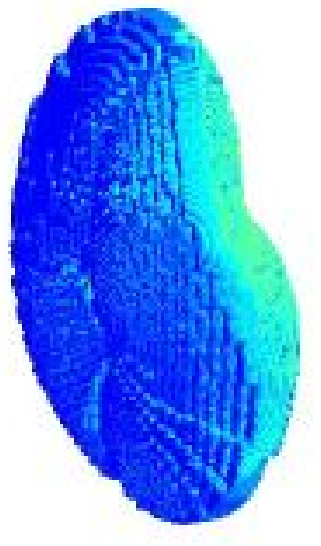,width=5cm}
			} &
		\subfigure[]{
			\label{mixed0_130_closed} 		
			\psfig{figure=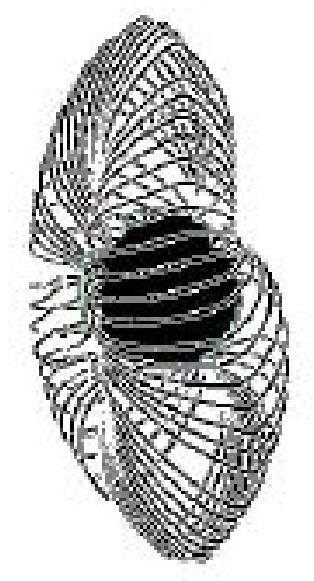,width=5cm}
			} &
		\subfigure[]{
			\label{mixed0_130_open} 
			\psfig{figure=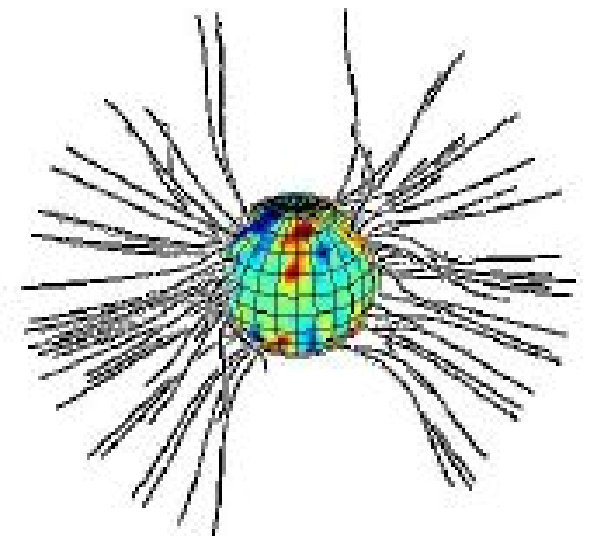,width=5cm}
			}  \\
	\end{tabular} 
\caption[]{Global field topology for a symmetric combination of the
1995 and 1996 surface maps where positive polarity open regions are at
the same longitude in each hemisphere (see also Fig.  \ref{icont}). 
The top three panels show a view from longitude 220$^{\circ}$ while
the bottom panels show a view from longitude 130$^{\circ}$.  Panels
\ref{mixed0_220_surf} and \ref{mixed0_130_surf} show the separator
surfaces that separate the closed and open field regions.  Panels
\ref{mixed0_220_closed} and \ref{mixed0_130_closed} show the closed
field lines just inside the separator and panels \ref{mixed0_220_open}
and \ref{mixed0_130_open} show the open field lines just outside it. 
In panels \ref{mixed0_220_open} and \ref{mixed0_130_open} the surface
radial map has been painted onto the stellar surface.  }
    \label{mixed0}
\end{figure*}	

\begin{figure*}
	\def\subfigtopskip{4pt}
	\def\subfigbottomskip{4pt}
	\def\subfigcapskip{2pt}
	\centering
	\begin{tabular}{ccc}
    	\subfigure[]{
			\label{mixed180_308_surf} 			
			\psfig{figure=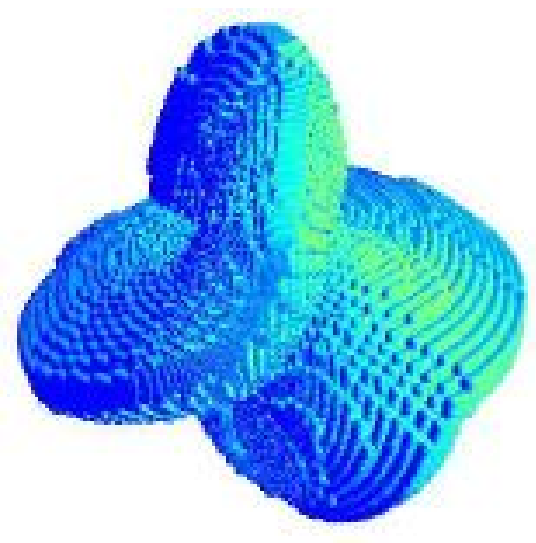,width=5.5cm}
			} &
		\subfigure[]{
			\label{308linesonly} 						
			\psfig{figure=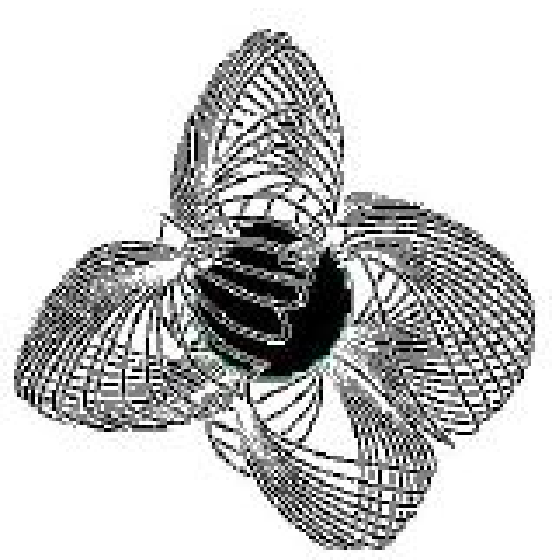,width=5cm}
			} &
		\subfigure[]{
			\label{308open} 
			\psfig{figure=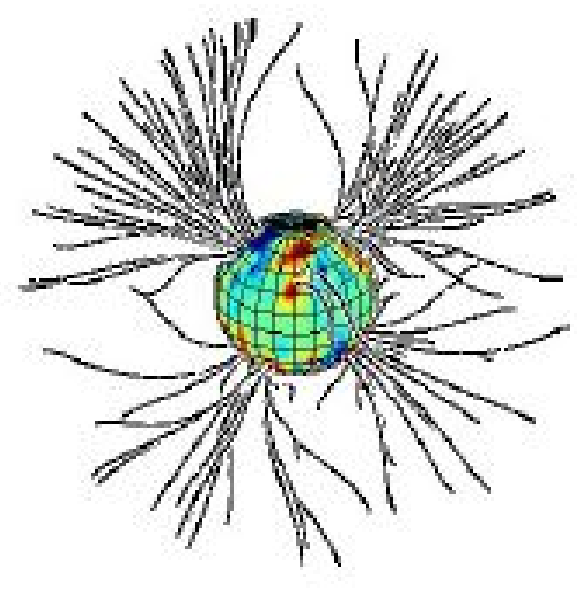,width=5cm}
			} \\
		\subfigure[]{
			\label{mixed180_128_surf} 					
			\psfig{figure=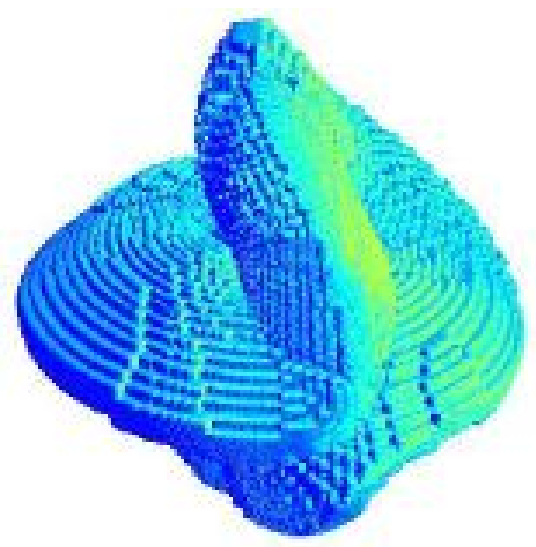,width=5.5cm}
			} &
		\subfigure[]{
			\label{128linesonly} 		
			\psfig{figure=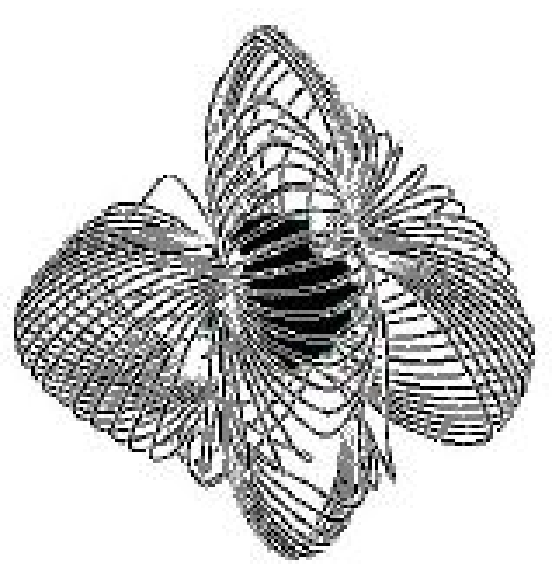,width=5cm}
			} &
		\subfigure[]{
			\label{128open} 
			\psfig{figure=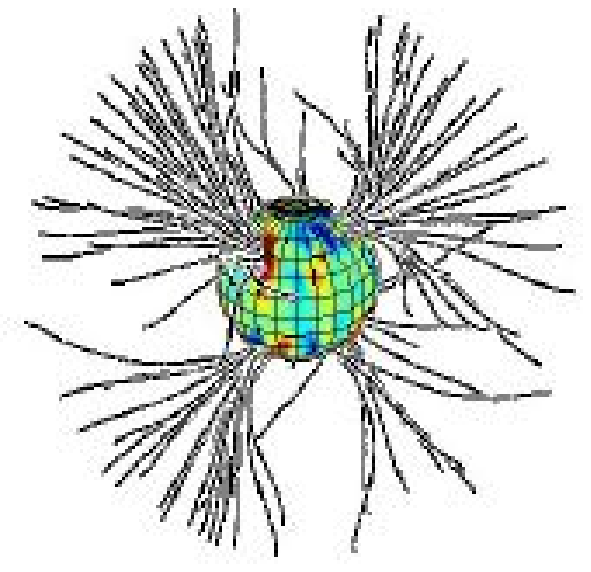,width=5cm}
			}  \\
	\end{tabular} 
\caption[]{As Fig. \ref{mixed0} except that the alignment of the two maps
is such that both hemispheres are anti-symmetric about the equator, 
with positive
polarity open regions 180$^{\circ}$ apart in longitude in the two
hemispheres (see also Fig. \ref{icont}). The top three
panels show a view from longitude 308$^{\circ}$ while the bottom panels
show a view from longitude 128$^{\circ}$.}
    \label{mixed180}
\end{figure*}	

Since the rotation axis of AB Dor is inclined at some 60$^{\circ}$ to
the observer, only one hemisphere can be imaged reliably.  The global
structure of the coronal field, however, depends on the way in which
field lines originating in the observable hemisphere connect to the
hidden hemisphere.  Close to the surface the effect of this missing
information is negligible as the small-scale fieldlines connect
locally to the surface.  The large-scale field is much more likely to
be affected.  While we have no way of determining the field structure
in the hidden hemisphere, we can assess the extent of its influence. 
We do this by creating an artificial surface map in which the 1995
data set forms one hemisphere and the 1996 map forms the other.  This
produces a global field structure where the two hemispheres have
similar large-scale field topologies.  It is of course possible that
the invisible hemisphere has a different structure to the one that we
can observe.  If, for example, the dynamo were to excite a mixture of
modes that are symmetric and antisymmetric about the equator then the
magnetic structure of the two hemispheres could be quite different. 
In the light of our limited information, however, we choose to make
the simplest assumption, that the lowest-order field in the two
hemispheres is either purely symmetric or antisymmetric about the
equator.  We obtain these two cases by selecting the alignment of the
maps for the two hemispheres.  In one case we choose a symmetric
alignment, such that the two positive regions of open field in each
hemisphere are at the same longitude.  In the other case we place them
180$^{\circ}$ of longitude apart (see Fig.  \ref{icont}).  In the
``symmetric'' case, we therefore have very extended coronal holes that
reach from one hemisphere into the other, while in the
``anti-symmetric'' case, the low-latitude field regions that were
originally open now connect across the equator to form closed field
regions.

In the symmetric case (Fig.  \ref{mixed0}), the closed corona is
confined to a torus that covers both poles and has an axis that
connects the longitudes of the open field regions.  The open field
regions extend across both hemispheres, reaching to fairly high
latitudes, but not to the poles.  In the anti-symmetric case (Fig. 
\ref{mixed180}), the closed corona is a complex surface that appears
to be the sum of two tori: one torus is similar to that for the
symmetric case and the other torus lies in the equatorial plane and
has its axis parallel to the rotation axis.  This second torus is
formed by field lines from low to intermediate latitudes in each
hemisphere connecting across the equator.

In both cases, the structure of the high-latitude field and the
longitudes of the coronal holes in the visible hemisphere are very
similar.  Indeed, they are very much the same as in the case where
there is no field added into the hidden hemisphere at all.  The main
difference is in the structure of the low to intermediate latitude
field.  In the symmetric case, any X-ray emission would have to come
from two distinct longitude bands (indeed, any prominences formed
would have to originate within these bands too).  This is in conflict
with the wide range of rotation phases at which prominences are
observed.  The antisymmetric case would allow prominence formation at
any longitude and also perhaps a greater X-ray emission measure, since
a greater volume of the corona is closed.  In the symmetric case, the
volume filling factor (i.e. the volume of the closed corona as a
fraction of the entire volume out to the source surface) is $0.24$,
whereas in the antisymmetric case it is $0.38$.

\section{The effect of flux hidden in the dark polar cap}

Doppler images of AB Dor consistently show that the polar regions
above latitude $\approx 70^{\circ}-80^{\circ}$ are dark.  The appearance of a
dark area suggests that the field in this region is strong enough to
inhibit convection, but Zeeman-Doppler imaging recovers little if any
field here.  This is principally because the Zeeman signal is
suppressed in areas of the surface that are dark.  We are therefore
unable to determine either the polarity or strength of this field (or
indeed to determine if it is of uniform or mixed polarity).

While a dark polar cap of mixed polarity would have limited effect on the
global field, the same is not true of a unipolar region.  In the
limiting case where this polar field dominates the field structure,
the closed corona would be in the form of a torus lying in the
equatorial plane and the closed field regions seen at the pole in Figs. 
\ref{mixed0} and \ref{mixed180} would be replaced by open field.  In order to
determine the extent to which such a polar field might affect the 
global field structure, we have added dipolar fields of different
strengths to the observed surface map for 1995.  The surfaces bounding
the closed corona are shown in Fig. {\ref{dipole}.  A dipole field
whose strength at the pole is only around 100G is sufficient to sweep
the polar regions clear of closed field lines.  Such a field could
easily pass undetected in the Zeeman-Doppler maps, but would be
insufficient to suppress convection.
\begin{figure*}
	\def\subfigtopskip{4pt}
	\def\subfigbottomskip{4pt}
	\def\subfigcapskip{2pt}
	\centering
	\begin{tabular}{ccc}
    	        \subfigure[]{
			\label{onehemis95_dip_p05_143} 			
			\psfig{figure=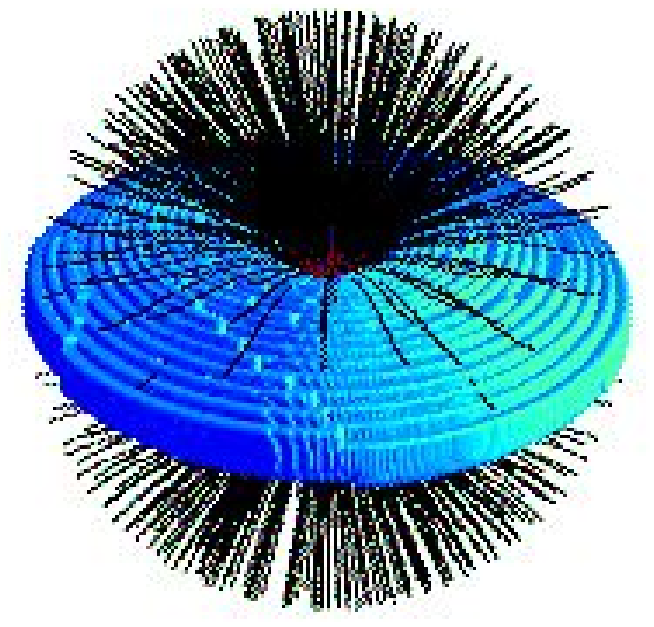,width=5cm}
			} &
		\subfigure[]{
			\label{onehemis95_dip_p01_143} 						
			\psfig{figure=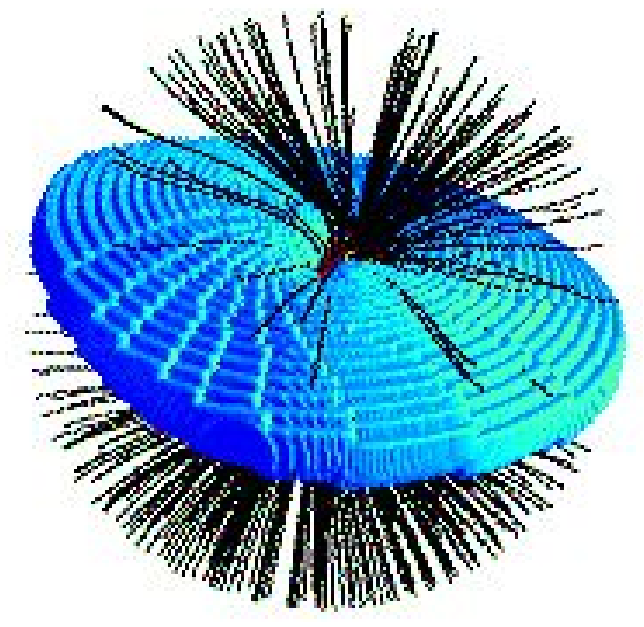,width=5cm}
			} &
		\subfigure[]{
			\label{onehemis95_dip_p005_143} 
			\psfig{figure=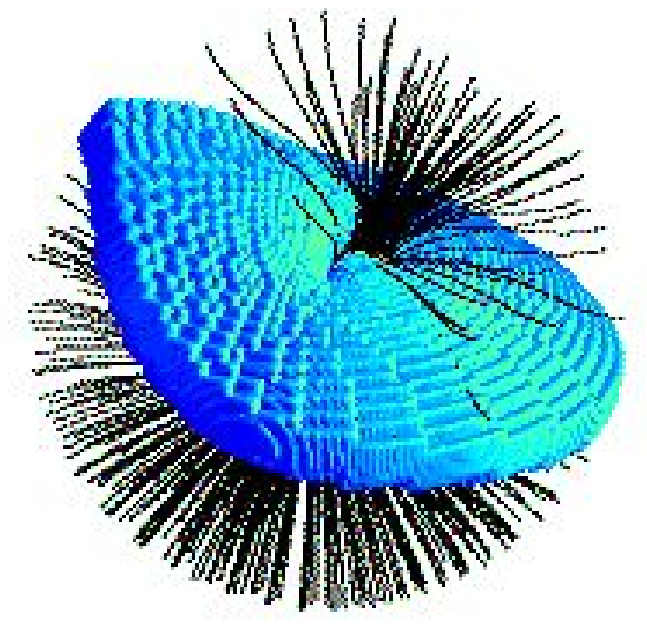,width=5cm}
			} \\
		\subfigure[]{
			\label{onehemis95_dip_p001_143} 					
			\psfig{figure=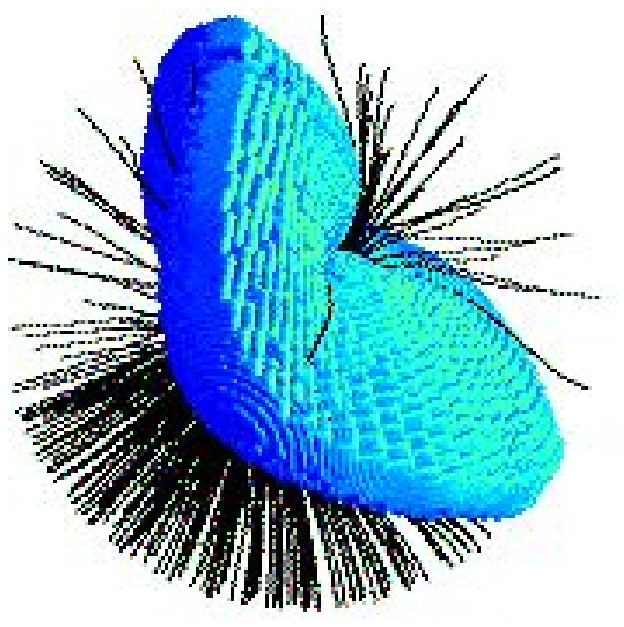,width=5cm}
			} &
		\subfigure[]{
			\label{onehemis95_dip_0_143} 		
			\psfig{figure=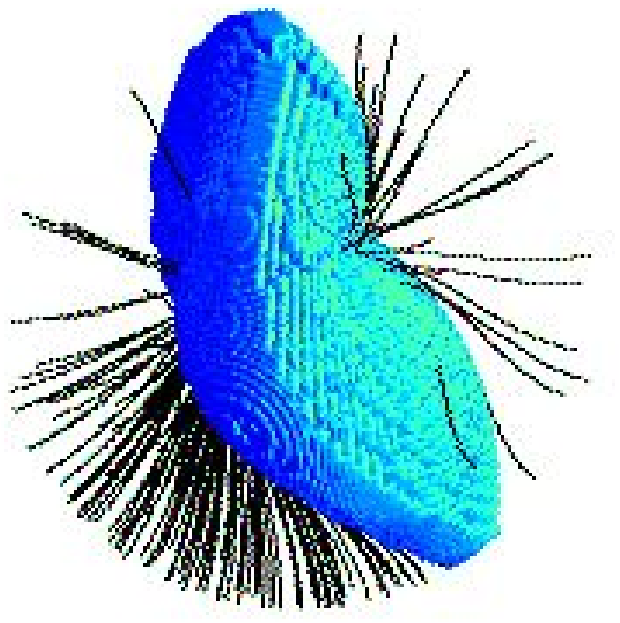,width=5cm}
			} &
		\subfigure[]{
			\label{onehemis95_dip_m001_143} 
			\psfig{figure=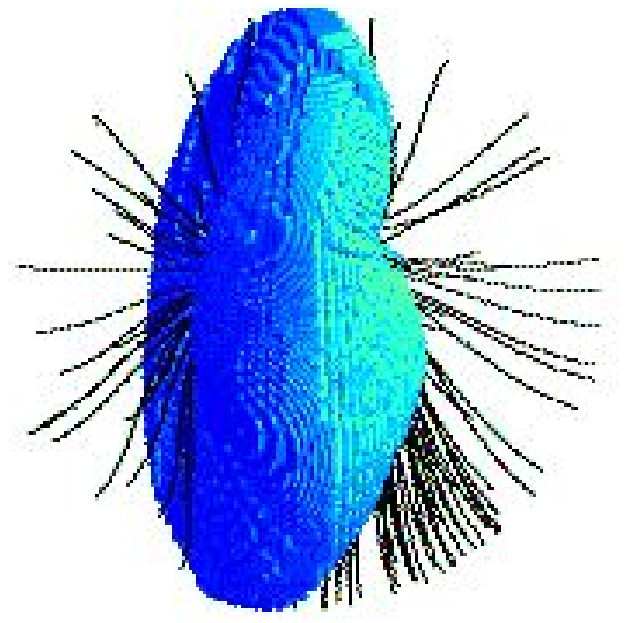,width=5cm}
			}  \\
		\subfigure[]{
			\label{onehemis95_dip_m005_143} 					
			\psfig{figure=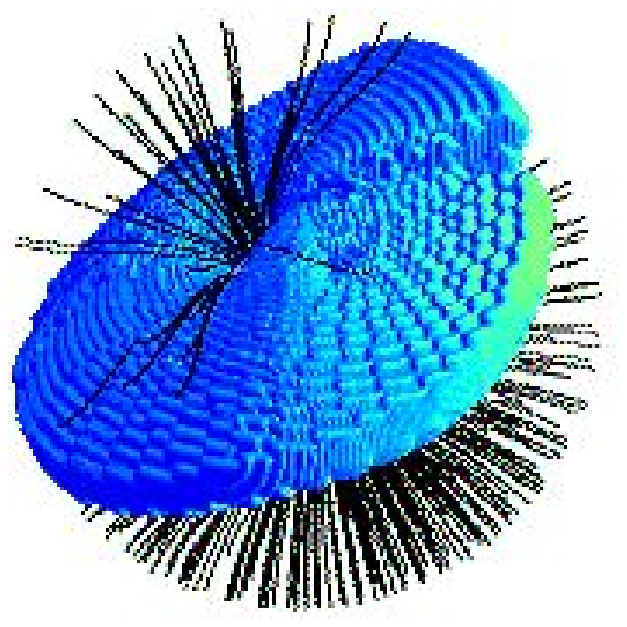,width=5cm}
			} &
		\subfigure[]{
			\label{onehemis95_dip_m01_143} 		
			\psfig{figure=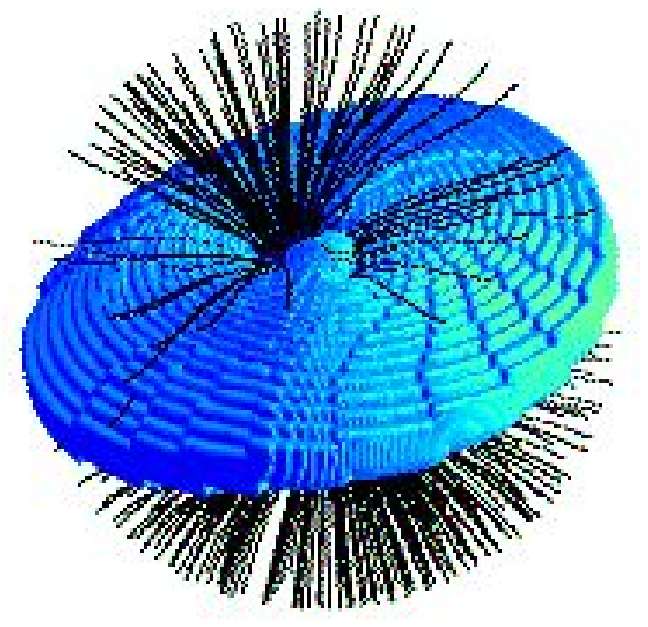,width=5cm}
			} &
		\subfigure[]{
			\label{onehemis95_dip_m05_143} 
			\psfig{figure=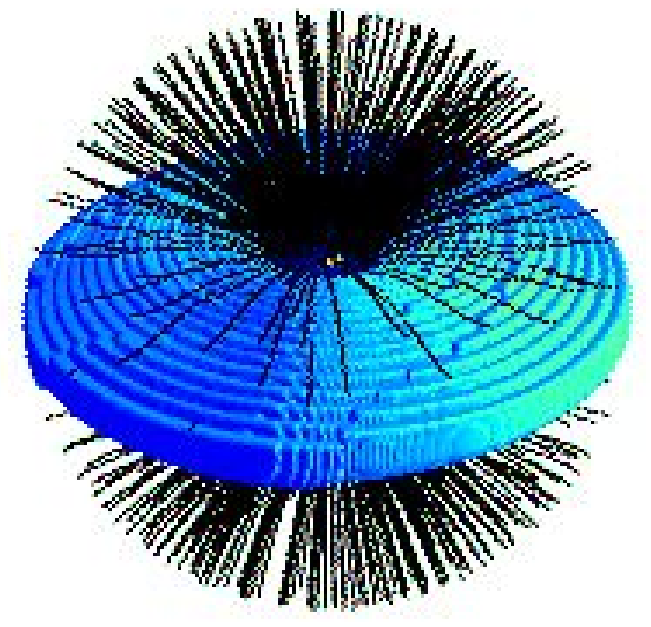,width=5cm}
			}  \\

	\end{tabular} 
\caption[]{Surfaces separating closed and open field regions for 
 different dipolar fields added to the surface maps. Field strengths added are: 
 \ref{onehemis95_dip_p05_143} 1000G, 
 \ref{onehemis95_dip_p01_143} 200G, 
 \ref{onehemis95_dip_p005_143} 100G, 
 \ref{onehemis95_dip_p001_143} 20G, 
 \ref{onehemis95_dip_0_143} 0G, 
 \ref{onehemis95_dip_m001_143} -20G, 
 \ref{onehemis95_dip_m005_143} -100G, 
 \ref{onehemis95_dip_m01_143} -200G, 
 \ref{onehemis95_dip_m05_143} -1000G. Sample open field lines are drawn in 
 each case.}
    \label{dipole}
\end{figure*}	

While the volume filling factor (i.e. the volume of the closed corona
as a fraction of the entire volume out to the source surface) of all
of these models is almost identical at $0.25$, what distinguishes them is
the orientation of the torus that forms the boundary of the closed
corona.  In particular, there are significant differences in the
latitudes at which the open field is found (and hence from which a
wind could escape).  The ratio of the open flux to the total flux is
in fact a maximum when the imposed polar field dominates over the
observed field.  It should be stressed that although we have
shown results for modelling the polar cap by the addition of a dipolar
field to the observed field, the addition of polar spots containing
the same flux as the dipole case gives qualitatively the same results.

\section{The limiting case of a polar field dominating the global 
topology}

The inevitable result of imposing a polar field that is strong enough
to suppress convection is that this field dominates on the largest
scales.  The global field topology is just that of a dipolar field
with a source surface imposed.  In particular, as can be seen from
\ref{onehemis95_dip_p05_143} or \ref{onehemis95_dip_m05_143}, the
polar regions are open and the polar hole extends down to a latitude
that depends on our choice of the source surface position.  This
result appears to be in conflict with the BeppoSAX observations of
\scite{maggio2000} that suggest the presence of closed loop structures
at high latitudes.

In order to explore the implications of such a polar field, we can 
examine the tractable case of a purely dipolar field. We therefore 
take the $l=1,m=0$ 
component of (\ref{psifull}) and impose the boundary conditions 
\begin{eqnarray}
    B_{r}(r=R_{\star}) & = & 2M\cos\theta / R^{3} \\
    B_{\theta}(r=R_{s}) & = & 0
\end{eqnarray}
where M is the dipole moment for a purely dipolar field and can be
defined as $M=B_{r}(r=R_{\star}, \theta=0) R_{\star}^{3}/2$.  These boundary
conditions give a ``pseudo-dipole'' field which can be written in
terms of a correction to the classical dipole field which allows for
the effect of the source surface:
   \begin{eqnarray}
       B_{r} & = & \frac{2M \cos \theta}{r^{3}} 
                 \left(
		     \frac{r^{3} +  2R_{s}^{3}}
		          {R_{\star}^{3} + 2R_{s}^{3}}
                 \right)    \\
       B_{\theta} & = & \frac{M \sin \theta}{r^{3}} 
                 \left(
		      \frac{-2r^{3} +  2R_{s}^{3}}
		      {R_{\star}^{3} + 2R_{s}^{3}}
		 \right).
   \label{components}
   \end{eqnarray}
The equation of a field line is $\sin^{2}\theta = Ar/(r^{3}+2R_{s}^{3})$.
The last closed field line passes through $\theta=\pi/2$ at $r=R_{s}$ 
and so has $A=3R_{s}^{2}$. It connects to the stellar surface at a 
co-latitude $\Theta_{0}$ where 
\begin{equation}
    \sin^{2}\Theta_{0} = \frac{3R_{s}^{2} R_{\star}}{R_{\star}^{3}+2R_{s}^{3}}.
\end{equation}
\begin{figure}
    
 \psfig{figure=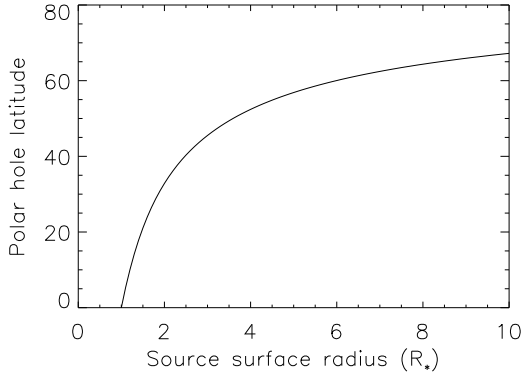,width=8cm} 
 
 \caption{Latitude of the boundary of the polar hole as a
 function of the source surface radius.  Beyond this spherical surface
 the field lines are forced to be open.  }

  \label{lastclosed}
 
\end{figure}
\begin{figure}
    
 \psfig{figure=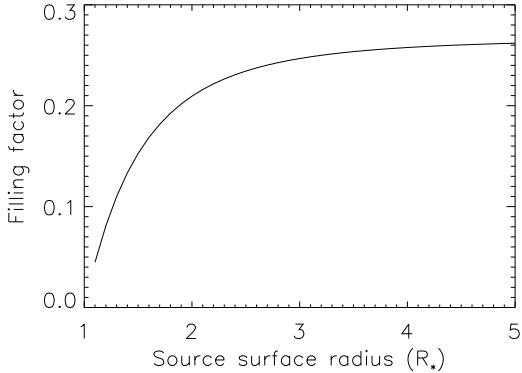,width=8cm} 
 
 \caption{Volume filling factor of the closed corona as  a 
 function of the source surface where the field lines are forced to be 
 open.}

  \label{fillfact}
 
\end{figure}

Fig \ref{lastclosed} shows how the latitude of this last closed field line 
varies with the imposed source surface radius $R_{s}$. For values of 
$R_{s}$ below about 3R$_{\star}$ this limiting latitude decreases rapidly, 
and the size of the polar hole increases accordingly. The source 
surface would have to be placed beyond 6$R_{\star}$ to push the polar 
hole above 60$^{\circ}$ latitude and to allow part of the closed 
field region to be uneclipsed, as in the Beppo-SAX flare 
\cite{maggio2000}. This would reduce the fraction of open flux, 
since the flux of open field through the stellar surface in 
the upper hemisphere is just $2 \pi R_{\star}^2 \int_{0}^{\Theta_{0}} 
B_{r}\sin \theta d \theta$. Hence the ratio of this open flux to the 
total flux through the upper hemisphere is simply $\sin^{2}\Theta_{0}$. 
As the source surface is moved further from the stellar surface, 
the fraction of the flux that is open decreases.

While the amount of open flux is important to the rate at which
angular momentum can be lost in a stellar wind, it is the volume of
the closed corona that is relevant to the amount of X-ray emission
that is produced.  For this pseudo-dipole field the volume filling factor can
be calculated by integrating the volume under the last closed field
line whose path $\theta=h(r)$ is defined by $\sin^{2}\theta = 
3R_{s}^{2}r/(r^{3}+2R_{s}^{3})$:
\begin{eqnarray}
    {\rm Volume} & = & 4 \pi \int_{\theta=h(r)}^{\theta=\pi /2} 
                       \int_{R_{\star}}^{R_{s}} r^{2} \sin \theta dr d\theta \\
          &  = & 4 \pi \int_{R_{\star}}^{R_{s}} r^{2}
	            \left(
		         1-\frac{3R_{s}^{2} r}{r^{3}+2R_{s}^{3}}
		    \right)^{1/2}   dr.
\end{eqnarray}
The dependence of this volume on the position of the source surface is
shown in Fig.  \ref{fillfact}.  As the source surface moves closer to
the star, the filling factor drops rapidly.  While the X-ray emission
measure also depends on the density, this shrinking of the available
volume will have an increasing effect as the source surface approaches
the stellar surface.  Jardine \& Unruh (1999) have shown that if rapid
rotation strips open the outer parts of stellar coronae then this
shrinking of the coronal volume with increasing rotation rate could
explain the observed saturation and supersaturation of the X-ray
emission \nocite{jardine99stripping}.

What is clear from Fig. \ref{lastclosed} is that in a star where the 
polar field dominates, the only way to have the corona extending to 
latitudes above about 60$^{\circ}$ is to have the source surface beyond 
about 6$R_{\star}$. This is consistent with the observations of 
prominences forming between 3 and 5 $R_{\star}$, but does raise the 
problem of containment. The equatorial co-rotation radius on AB Dor is 
at some 2.7R$_{\star}$ from the rotation axis. Beyond this point, for 
an isothermal corona, the density and pressure of the corona start to 
rise. The magnetic pressure falls with height, however, and so 
inevitably at some point the plasma pressure will exceed the 
magnetic pressure. The plasma will be able to distort and ultimately 
open up the field lines. Our imposition of a source surface is a crude 
way of modelling this process. If we place the source surface far beyond 
the co-rotation radius, an implausibly strong field is required to 
confine the plasma.

We can quantify this problem by determining the plasma pressure as a 
function of height for the case where the polar field dominates. If we 
impose hydrostatic equilibrium along a field line, then the pressure 
is simply 
\begin{equation}
    p=p_{0} \exp{ \left( \frac{m}{kT}\oint_{s}g_{s}ds \right) }.
\end{equation}
Here T is the temperature, $g_{s}$ is the component 
of the effective gravity along the 
field line, i.e. $g_{s} =( {\bf g.B})/|{\bf B}|$ and 
\begin{equation}
g(r,\theta) = \left( -GM_{\star}/r^{2} + 
                     \omega^{2}r\sin^{2}\theta,
		     \omega^{2}r\sin\theta\cos\theta 
             \right), 
\end{equation}
where $\omega$ is the angular velocity. Using (\ref{components}) we 
can write this as
\begin{equation}
  p(r) = p_{0} \exp{  \left( -\Phi_{g}\left(  1-\frac{R_{\star}}{r} \right) 
                + \Phi_{c} \frac{(r/R_{\star})^{3}-1}{(r/R_{s})^{3}+2} C
		    \right)  }
\end{equation}
where  $\Phi_{g}$ 
and $\Phi_{c}$ are the surface ratios of the gravitational and 
centrifugal energies to the thermal energies, i.e. 
\begin{eqnarray*}
    \Phi_{g} &= & \frac{GM_{\star}/R_{\star}}{kT/m} \\
    \Phi_{c} & = & \frac{\omega^{2}R_{\star}^{2}/2}{kT/m}.
\end{eqnarray*} 
The constant C varies from one field line to the next. If we 
focus on the equatorial plane, where each field line has its maximum 
extent $r_{m}$, then C is given by
\begin{equation}
    C = \frac{2R_{\star}(r_{m}^{3} + 2 R_{s}^{3})}{r_{m}(R_{\star}^{3}+2 R_{s}^{3})}.
\end{equation} 

The variation with height of the magnetic pressure is more 
straightforward. Looking just at the equatorial plane, we find that
\begin{equation}
   |B| =\frac{2M}{r^{3}} \frac{(-r^{3}+R_{s}^{3})}{(R_{\star}^{3}+2R_{s}^{3})}.
\end{equation}

\begin{figure}
      
 \psfig{figure=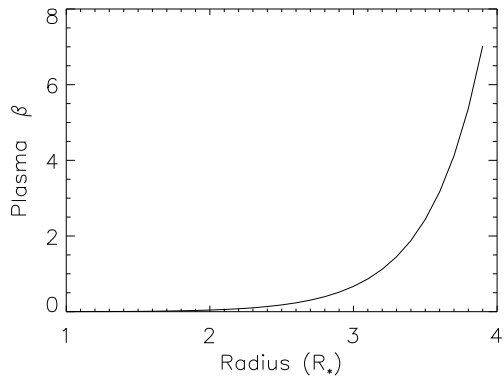,width=8cm} 
 
 \caption{The ratio $\beta$ of plasma to magnetic pressure along a
 field line as a function of radius. The field
 lines are forced to be open at a radius $R_{s}=6R_{\star}$.}

  \label{beta}
 
\end{figure}

If we now take the ratio $\beta$ of the plasma pressure to the
magnetic pressure ($B^{2}/2\mu$) we find that it rises rapidly with
distance above the stellar surface.  If, as in Fig.  \ref{beta} we put
the source surface at 6$R_{\star}$, we find that the ratio $\beta$ reaches 
unity
at around $r=3.3R_{\star}$ for a base pressure of 1Pa, a temperature of
10$^{7}$K and a field strength at the pole of 1kG. This pressure is
consistent with the results of \scite{maggio2000}.
The height at which $\beta=1$ is in fact fairly
insensitive to our choice of parameters.  Placing the source surface
further out (say to 10$R_{\star}$) or dropping the coronal temperature
from $10^{7}$K to $10^{6}$K moves this point out closer to
4R$_{\star}$, but in either case, the plasma pressure is one or two
orders of magnitude greater than the magnetic pressure before the
source surface is reached. Even with a base pressure as low as 0.01Pa 
with the source surface at 10$R_{\star}$, the field lines on which  
$\beta>1$ do not extend above latitude 60$^{\circ}$. 

\section{Conclusions}

We have used Zeeman-Doppler images of the surface magnetic field of AB
Dor to extrapolate the coronal magnetic field.  This field has a clear
non-axisymetric structure that becomes very apparent upon tracing the
open field lines that fill much of the coronal volume.  These open
field lines originate in two opposite-polarity mid-latitude regions,
separated by about 180$^{\circ}$ of longitude.  The surface separating
the closed and open field regions forms a torus that passes over the
visible pole.  We have tried to allow for the two regions in which
flux could be missing from these maps: both in the unobservable
hemisphere and in the dark polar cap.  Allowing for field in the
unobservable hemisphere changes the field line connections at low
latitudes, but the high-latitude field topology is largely unaffected. 
If we allow for the flux that may be concealed in the dark polar cap,
we find that the polar field lines become open with the addition of a
polar field so weak that it would not suppress convection sufficiently
to cause the polar regions to be dark.  If we impose a polar field
strong enough to give a dark polar cap, we find that the large scale
field is similar to that of a dipole with a source surface imposed.

This ``pseudo-dipole'' field does not however explain the BeppoSAX
flare observations which imply high latitude closed loops.  The
boundary of the polar hole reaches down to latitudes well below that
where observations suggest the footpoints of the flaring loop must
have been located.  The only way to reduce the size of the polar hole
sufficiently is to move the source surface out to beyond 6$R_{\star}$. 
This, however, requires the pressure in the tops of the largest loops
to be much greater than the magnetic pressure.  The only reasonable
solution seems to be to allow the polar flux to be of mixed polarity. 
This could be achieved by having the dark polar cap composed of many
smaller spots of mixed polarity, as in the flux-emergence models of
Sch\"ussler \etal (1996), \nocite{schussler96} or by having a smaller
polar spot surrounded by a ring of opposite polarity as in the models
of Schrijver \& Title (2001).  \nocite{schrijver01} Field lines in the
polar spot will connect preferentially to the surrounding
opposite-polarity ring, forming closed field regions at high
latitudes.

If indeed young magnetically active stars do have mixed-polarity
regions in their dark polar caps, there are implications for models of
pre-main sequence stars.  Many earlier models for the structure and
formation of accretion disks and jets have rested upon the assumption
that the star itself has a dipole-like field
\cite{shu94,lovelace95,miller97}.  More recently, Agapitou \&
Papaloizou (2000) and Terquem \& Papaloizou (2000) have shown that
departures from a potential field or a mis-alignment of the dipole
field with the stellar rotation axis (i.e. a non-axisymmetric
structure) can influence the angular momentum transfer between the
star and the disk or cause an observable warping of the disk. 
\nocite{agapitou2000,terquem2000} Angular momentum loss in a stellar 
wind 
is also sensitively dependent on the surface positions from which open 
field lines escape \cite{solanki97}. In Fig. \ref{dipole} the number 
of open field lines drawn in each panel is directly proportional to 
the surface area of open field lines. The nature of the field in the 
polar cap clearly has a significant influence on the structure of the open 
field and hence potentially on the spin-down rate of young stars. 
Our Zeeman-Doppler images imply that
allowing for departures from dipolar fields is certainly
justified.

\section*{ACKNOWLEDGEMENTS}
The authors would like to thank Dr. A. van Ballegooijen for 
allowing us to use his code to calculate the potential magnetic field and 
Dr. D. Mackay for help in implementing it. 
 
\bibliographystyle{mn}
\bibliography{iau_journals,master,ownrefs,mmjpapers}

\begin{thebibliography}{{{Solanki, S.K.}, {Motamen, S.} \& {Keppens,
  R.}}{1997}}

\bibitem[\protect\citefmt{{Agapitou} \& {Papaloizou}}{2000}]{agapitou2000}
{Agapitou}~V., {Papaloizou}~J., 2000, MNRAS, 317, 273

\bibitem[\protect\citefmt{{Altschuler} \& {Newkirk, Jr.}}{1969}]{altschuler69}
{Altschuler}~M.~D., {Newkirk, Jr.}~G., 1969, Solar~Phys., 9, 131

\bibitem[\protect\citefmt{{Amado} {\rm et~al.}}{2001}]{amado01}
{Amado}~P.~J., {Cutispoto}~G., {Lanza}~A.~F., {Rodon{\` o}}~M., 2001, in ASP
  Conf. Ser. 223: 11th Cambridge Workshop on Cool Stars, Stellar Systems and
  the Sun.
\newblock p.~895+

\bibitem[\protect\citefmt{Collier~Cameron \& Robinson}{1989a}]{cameron89cloud}
Collier~Cameron~A., Robinson~R.~D., 1989a, MNRAS, 236, 57

\bibitem[\protect\citefmt{Collier~Cameron \& Robinson}{1989b}]{cameron89eject}
Collier~Cameron~A., Robinson~R.~D., 1989b, MNRAS, 238, 657

\bibitem[\protect\citefmt{Donati \& Collier~Cameron}{1997}]{donati97abdor95}
Donati~J.-F., Collier~Cameron~A., 1997, MNRAS, 291, 1

\bibitem[\protect\citefmt{Donati {\rm et~al.}}{1999}]{donati99abdor96}
Donati~J.-F., Collier~Cameron~A., Hussain~G., Semel~M., 1999, MNRAS, 302, 437

\bibitem[\protect\citefmt{{Jardine} \& {Unruh}}{1999}]{jardine99stripping}
{Jardine}~M., {Unruh}~Y., 1999, A\&A, 346, 883

\bibitem[\protect\citefmt{{K\"urster} {\rm et~al.}}{1997}]{kurster97}
{K\"urster}~M., {Schmitt}~J., {Cutispoto}~G., {Dennerl}~K., 1997, A\&A, 320,
  831

\bibitem[\protect\citefmt{{Lim} {\rm et~al.}}{1994}]{lim94}
{Lim}~J., {White}~S., {Nelson}~G., {Benz}~A., 1994, ApJ, 430, 332

\bibitem[\protect\citefmt{{Lovelace}, {Romanova} \&
  {Bisnovatyi-Kogan}}{1995}]{lovelace95}
{Lovelace}~R.~V.~E., {Romanova}~M.~M., {Bisnovatyi-Kogan}~G.~S., 1995, MNRAS, 275,
  244

\bibitem[\protect\citefmt{{Maggio} {\rm et~al.}}{2000}]{maggio2000}
{Maggio}~A., {Pallavicini}~R., {Reale}~F., {Tagliaferri}~G., 2000, A\&A, 356,
  627

\bibitem[\protect\citefmt{{Miller} \& {Stone}}{1997}]{miller97}
{Miller}~K.~A., {Stone}~J.~M., 1997, ApJ, 489, 890+

\bibitem[\protect\citefmt{{Pointer} {\rm et~al.}}{2002}]{pointer01evol}
{Pointer}~G.~R., {Jardine}~M., {Collier Cameron}~A., {Donati}~J.-F., 2002,
  MNRAS, 330, 160+

\bibitem[\protect\citefmt{{Schrijver} \& {Title}}{2001}]{schrijver01}
{Schrijver}~C.~J., {Title}~A.~M., 2001, ApJ, 551, 1099

\bibitem[\protect\citefmt{{Sch\"ussler} {\rm et~al.}}{1996}]{schussler96}
{Sch\"ussler}~M., {Caligari}~P., {Ferriz-Mas}~A., {Solanki}~S., {Stix}~M.,
  1996, A\&A, 314, 503

\bibitem[\protect\citefmt{{Shu} {\rm et~al.}}{1994}]{shu94}
{Shu}~F., {Najita}~J., {Ostriker}~E., {Wilkin}~F., {Ruden}~S., {Lizano}~S.,
  1994, ApJ, 429, 781

\bibitem[\protect\citefmt{{Solanki, S.K.}, {Motamen, S.} \& {Keppens,
  R.}}{1997}]{solanki97}
{Solanki, S.K.}, {Motamen, S.}, {Keppens, R.}, 1997, A\&A, 324, 943

\bibitem[\protect\citefmt{{Terquem} \& {Papaloizou}}{2000}]{terquem2000}
{Terquem}~C., {Papaloizou}~J., 2000, A\&A, , 1031

\bibitem[\protect\citefmt{van Ballegooijen, Cartledge \&
  Priest}{1998}]{vanballegooijen98}
van Ballegooijen~A., Cartledge~N., Priest~E., 1998, ApJ, 501, 866

\bibitem[\protect\citefmt{{Vilhu} {\rm et~al.}}{2001}]{vilhu2001}
{Vilhu}~O., {Mulhi}~P., {Mewe}~R., {Hakala}~P., 2001, A\&A, 375, 492

\end{thebibliography}

\end{document}